\documentstyle[12pt,epsf]{article}

\begin{document}
\title{
  \begin{flushright}
    \large UT-752
  \end{flushright}
  \vspace{5ex}
  Dynamical Supersymmetry Breaking without Messenger Gauge
  Interactions}
\author{T. Hotta, Izawa K.-I. and T. Yanagida\\
  \\  Department of Physics, University of Tokyo \\
  Bunkyo-ku, Tokyo 113, Japan}
\date{\today}
\maketitle

\begin{abstract}
We investigate low-energy models of supersymmetry (SUSY) breaking by
means of vector-like gauge theories for dynamical SUSY breaking.
It is not necessary to introduce messenger gauge interactions utilized
so far to mediate the SUSY breaking to the standard-model sector,
which reduces complication in the model building.
We also consider various other ways of SUSY-breaking transmission.
\end{abstract}

\newpage

\section{Introduction}

Supersymmetry (SUSY) is a plausible candidate as an origin of the
hierarchy between the weak scale and higher scales, say the grand
unification (GUT) and/or the gravitational scales.
It must be broken at low energy so that the observed particles do not
have their superpartners with the same masses.
We consider dynamical SUSY breaking (DSB) as the mechanism of SUSY
breaking at low energy since DSB gives a complete solution to the
hierarchy problem --- it can induce a very tiny scale compared with
the gravitational scale without any fine-tunings of parameters in the
theory.
In particular, low-energy models of DSB (so-called visible sector
models) are attractive also in that they can give a natural solution
to the flavor changing neutral current (FCNC) problem through the mass
degeneracy of the squarks and sleptons \cite{dine}.

There are two methods to mediate the low-energy SUSY breaking in the
DSB sector to the SUSY standard-model sector.
One is to identify a global symmetry in the DSB sector as gauge groups
in the standard model.%
\footnote{We note that this is easily realized in the SP$(N)$
  vector-like gauge theories of DSB proposed in Ref.\cite{izawa},
  since the anomaly-free condition for the standard-model gauge
  interactions is manifestly satisfied without introducing additional
  fields.}
This approach, however, has difficulty that the SU$(3)_C$ gauge
coupling constant tends to blow up at too low an energy scale.
The other is to introduce a messenger sector which couples to the both
sectors.

The phenomenological models with low-energy DSB proposed so far
\cite{dine} share the common structures: DSB is realized in chiral
gauge theories and SUSY-breaking effects are mediated to the SUSY
standard-model sector by an additional messenger gauge interaction
which is introduced only for that purpose.
The additional gauge interaction is needed since the DSB and the
standard-model sectors are both chiral.

In this paper we investigate a model with a vector-like gauge theory
as the SUSY-breaking sector. We adopt vector-like models of DSB
proposed recently in Ref.\cite{izawa}.
We see that it is not necessary to introduce the messenger gauge
interaction, whose coupling threatens to blow up.
Note that this may largely reduce complication in the model building.
We also consider various other ways of SUSY-breaking transmission.

\section{The SUSY-Breaking Sector}

Let us consider a SUSY SU(2) gauge theory with four doublet chiral
superfields $Q_i$ and six singlet ones $Z^{ij} = -Z^{ji}$.
Here $i$ and $j$ denote the flavor indices ($i, j = 1, \cdots, 4$).
Without a superpotential, this model has a flavor SU(4)$_F$ symmetry.

The tree-level superpotential of the model \cite{izawa} is given by
\begin{equation}
  W_{tree} = \lambda_{ij}^{kl} Z^{ij} Q_k Q_l,
\end{equation}
where $\lambda_{ij}^{kl}$ denote generic coupling constants with
$\lambda_{ij}^{kl} = -\lambda_{ji}^{kl} = -\lambda_{ij}^{lk}$.
SUSY remains unbroken perturbatively in this model.

The exact effective superpotential of the model, which takes into
account the full nonperturbative effects, may be written in terms of
gauge-invariant low-energy degrees of freedom
\begin{equation}
  V_{ij} = -V_{ji} \sim Q_i Q_j
\end{equation}
as follows:
\begin{equation}
  W_{eff} = X({\rm Pf} \, V_{ij} - \Lambda^4) + \lambda_{ij}^{kl}
  Z^{ij} V_{kl},
\end{equation}
where $X$ is an additional chiral superfield, ${\rm Pf} \, V_{ij}$
denotes the Pfaffian of the antisymmetric matrix $V_{ij}$, and
$\Lambda$ is a dynamical scale of the SU(2) gauge interaction.
This is none other than a superpotential of the O'Raifeartaigh type.
Namely, this effective superpotential yields conditions for SUSY vacua
\begin{equation}
  {\rm Pf} \, V_{ij} = \Lambda^4, \quad \lambda_{ij}^{kl} V_{kl} = 0,
\end{equation}
which are incompatible as long as $\Lambda \neq 0$.
Therefore we conclude that SUSY is dynamically broken in this model
\cite{izawa}.

Let us impose a flavor SP(4)$_F$ symmetry on the above model to make
our analysis simpler, where we adopt the notation ${\rm SP}(4)_F
\subset {\rm SU}(4)_F$.
Then the effective superpotential can be written as
\begin{equation}
  \label{dsbPot}
  W_{eff} = X(V^2 + V_a V_a - \Lambda^4) + \lambda_Z Z V + \lambda Z^a 
  V_a,
\end{equation}
where $V$ and $Z$ are singlets and $V_a$ and $Z^a$ are
five-dimensional representations of SP(4)$_F$, respectively, in
$V_{ij}$ and $Z^{ij}$, which constitute six-dimensional
representations of SU(4)$_F$.
Here $a = 1, \cdots, 5$ and $\lambda_Z$ and $\lambda$ denote coupling
constants which are taken to be positive.

When the coupling $\lambda_Z$ is small, the effective superpotential
$W_{eff}$ implies that we obtain the following vacuum expectation
values:
\begin{equation}
  \label{vevs}
  \langle V \rangle \simeq \Lambda^2, \quad \langle V_a \rangle \simeq 
  0.
\end{equation}
Then the low-energy effective superpotential may be approximated by%
\footnote{The fields $Z^a$ and $V_a$ form massive multiplets with mass 
  of order $\lambda \Lambda$ and they are integrated out.}
\begin{equation}
  \label{effPot}
  W_{eff} \simeq \lambda_Z \Lambda^2 Z.
\end{equation}

On the other hand, the effective K\"ahler potential is expected to
take a form
\begin{equation}
  \label{kahler}
  K = ZZ^* - \frac{\eta}{4 \Lambda^2} \lambda_Z^4 (ZZ^*)^2 + \cdots,
\end{equation}
where $\eta$ is a real constant of order one.

Then the effective potential of the scalar $Z$ (with the same notation 
as the superfield) is given by
\begin{equation}
  V_{eff} \simeq \lambda_Z^2 \Lambda^4 (1 + \frac{\eta}{\Lambda^2}
  \lambda_Z^4 ZZ^* + \cdots).
\end{equation}
When $\eta > 0$, this leads to $\langle Z \rangle = 0$.
Otherwise, we suspect that $\langle Z \rangle$ is of order
$\lambda_Z^{-1} \Lambda$ since the effective potential is expected to
be lifted in the region far from the origin ($Z = 0$), as is the case
for the O'Raifeartaigh model \cite{fischler}.
In any event, the DSB scale is given by $F_Z \simeq \lambda_Z
\Lambda^2$, where $F_Z$ denotes the $F$ component of the superfield
$Z$, and the mass $m_Z$ of the scalar $Z$ is seen to be of order
$\sqrt{\eta} \lambda_Z^3 \Lambda$ except for the potential massless
$R$-axion.

We henceforth assume $\eta > 0$ for definiteness.

\section{The Messenger Sector}

The messenger sector consists of chiral superfields
$Y, d, \bar{d}, l, \bar{l}$, which are all singlets under the strong
SU(2) and the global SP(4)$_F$.
$Y$ is also a singlet under the standard-model gauge group.
As for $d, \bar{l}$ and $\bar{d}, l$ we tentatively assume that they
transform as the down quark, the anti-lepton doublet and their
antiparticles, respectively.
The interactions between the DSB and the messenger sectors are
described by a superpotential
\begin{equation}
  \label{intPot}
  W_{int} = \frac{\lambda_Y}{2} \epsilon_{\alpha \beta} \left
  ( Q_1^\alpha Q_2^\beta + Q_3^\alpha Q_4^\beta \right) Y
  - \frac{f}{3} Y^3 + (k_1 \bar{d} d + k_2 \bar{l} l) Y, 
\end{equation}
where $\alpha$ and $\beta$ denote SU(2) gauge indices and the
couplings $\lambda_Y$, $f$, $k_1$ and $k_2$ are taken to be positive.
In the following, we set $k \equiv k_1 = k_2$ for simplicity, which
approximately holds in SUSY-GUT's.

In view of Eqs.(\ref{vevs}) and (\ref{effPot}), the full effective
superpotential of the DSB and messenger sectors is obtained as
\begin{equation}
  \label{messPot}
  W_{eff} \simeq \lambda_Z \Lambda^2 Z + \lambda_Y \Lambda^2 Y
  - \frac{f}{3} Y^3 + k ( \bar{d} d + \bar{l} l) Y ,
\end{equation}
where we have used
\begin{equation}
  \label{SPsinglet}
  V \sim \frac{1}{2} \epsilon_{\alpha \beta} (Q_1^\alpha Q_2^\beta
  + Q_3^\alpha Q_4^\beta).
\end{equation}

\begin{figure}[tbp]
  \begin{center}
    \leavevmode
    \epsfbox{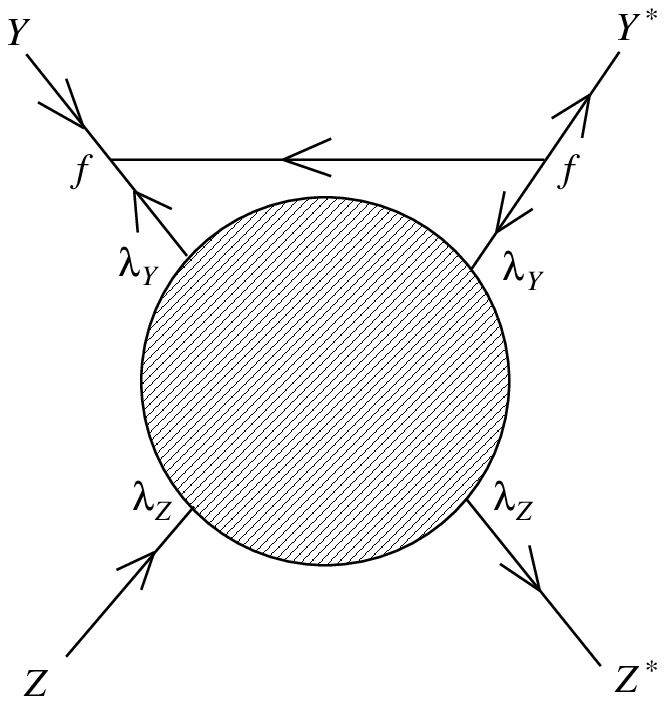}
  \end{center}
  \caption{}
  \label{fig:1}
\end{figure}

The K\"ahler potential of $Z$ and $Y$ is expected to have the 
following form:
\begin{equation}
 \label{Kahler}
  K = Z Z^* + Y Y^*
  - \frac{\eta}{4 \Lambda^2}|\lambda_Z Z + \lambda_Y Y|^4
  - \frac{\delta}{\Lambda^2} \frac{f^2}{16 \pi^2}
  \lambda_Y^2 |\lambda_Z Z + \lambda_Y Y|^2 \, Y Y^* + \cdots ,
\end{equation}
where $\delta$ is a real constant of order one.
We notice that non-anomalous $R$ and discrete symmetries with the
coupling $f$ as an external field%
\footnote{The $R$ charges of $Q, Z, Y, f$ read $0, 2, 2, -4$,
  respectively. 
  The discrete symmetry amounts to a transformation $Q \rightarrow
  iQ$, $Z \rightarrow -Z$, $Y \rightarrow -Y$, $f \rightarrow -f$.} 
may be utilized to see that there are no trilinear terms such as
$ZYY^*$ in the above K\"ahler potential.
As we will see below, we obtain vanishing $F$ component of $Y$,
$\langle F_Y \rangle = 0$, in a limit $f \rightarrow 0$.
In that case, the SUSY breaking is not transmitted to the
standard-model sector by means of the singlet $Y$.
Typical Feynman diagrams generating $f$-dependent corrections to the
K\"ahler potential in Eq.(\ref{Kahler}) are shown in Fig.\ref{fig:1}.

We obtain an effective potential from Eqs.(\ref{messPot}) and
(\ref{Kahler})
\begin{equation}
  \label{fullPot}
  \begin{array}{rl}
    V_{eff} \simeq & 
    \lambda_Z^2 \Lambda^4
    + |\lambda_Y \Lambda^2 - f Y^2 + k ( \bar{d} d + \bar{l} l)|^2
    + |k \bar{d} Y|^2 + |k d Y|^2 + |k \bar{l} Y|^2 + |k l Y|^2 \\
    \noalign{\vskip 1.0ex}
    & + \eta \lambda_Z^6 \Lambda^2 Z Z^*
    + \eta \lambda_Z^5 \lambda_Y \Lambda^2 ( Z Y^* + Z^* Y )
    + \eta \lambda_Z^4 \lambda_Y^2 \Lambda^2 Y Y^* \\
    \noalign{\vskip 0.5ex}
    & \displaystyle + \delta \frac{f^2}{16 \pi^2} 
    \lambda_Z^4 \lambda_Y^2 \Lambda^2 Y Y^*, \\
  \end{array}
\end{equation}
where we have taken into account only the leading corrections in the
coupling $\lambda_Y$.

We restrict ourselves to the vacua with $\langle d \rangle = \langle
\bar{d} \rangle = \langle l \rangle = \langle \bar{l} \rangle = 0$ in
the following consideration since, otherwise, the standard-model gauge 
group would be broken by their vacuum expectation values.
Notice that the above vacua is realized when the coupling $k$ is large 
enough.
Then the potential Eq.(\ref{fullPot}) yields a vacuum given by 
\begin{equation}
  \langle F_Y \rangle 
  \simeq \langle \lambda_Y \Lambda^2 - f Y^2 \rangle
  \simeq \delta \frac{f^2}{16 \pi^2} 
  \frac{\lambda_Z^4 \lambda_Y^2}{2 f} \Lambda^2, \quad
  \langle Y^2 \rangle \simeq \frac{\lambda_Y}{f} \Lambda^2 .
\end{equation}
The messenger quarks and leptons have masses $m_{d. l}$ of order
$\Lambda$ in this vacuum:
\begin{equation}
  m_{d. l} \simeq k \langle Y \rangle = k \sqrt{\frac{\lambda_Y}{f}}
  \Lambda.
\end{equation}
Since $\langle F_Y \rangle$ and $\langle Y \rangle $ are
non-vanishing, the SUSY breaking is mediated to the standard-model
sector as seen in the next section.

\section{The Standard-Model Sector}
\label{mu}

The standard-model gauginos obtain their masses from radiative
corrections through loops of the messenger quarks and leptons
\cite{dine}:
\begin{equation}
  \label{gauginomass}
  m_{\tilde{g_i}}
  \simeq \frac{g_i^2}{16 \pi^2} k^2 \frac{\langle Y \rangle \langle 
    F_Y \rangle}{m_{d, l}^2}
  \simeq \frac{\alpha_i}{4 \pi} \frac{\langle F_Y \rangle}{\langle Y
    \rangle}
  \simeq \frac{\alpha_i}{4 \pi}
  \delta \frac{f^2}{16 \pi^2}
  \frac{\lambda_Z^4 \lambda_Y^\frac{3}{2}}{2 f^\frac{1}{2}} \Lambda,
\end{equation}
where $\alpha_i = g_i^2/4\pi$ denote the standard-model gauge
couplings.

The masses of the squarks and sleptons are given by \cite{dine}
\begin{equation}
  \label{scalarM}
  \tilde{m}^2 \simeq 2(C_3 \, m_{gluino}^2 + C_2 \, m_{wino}^2
  + \frac{3}{5} \, C_1 \, m_{bino}^2) ,
\end{equation}
where $C_3 = 0$ or $4/3$ and $C_2 = 0$ or $3/4$, whose non-vanishing
values are the quadratic Casimir invariants for the fundamental
representations of SU(3)$_C$ and SU(2)$_L$, respectively, and $C_1$
denotes the corresponding hypercharges squared.
Namely, the squarks and sleptons have the masses of the same order as
the gaugino masses.
Notice here that the squarks and sleptons with the same quantum
numbers of the standard-model gauge group degenerate in mass, which
leads to the suppression of FCNC as noted in the Introduction.

We now proceed to consider phenomenological constraints on the
parameters in the potential Eq.(\ref{fullPot}).

First, the gluino mass should be ($10^2 - 10^3$) GeV to maintain the
weak scale of order $10^2$ GeV.
This is because growth of the gluino mass increases the stop mass
(see Eq.(\ref{scalarM})) and the large stop mass raises the Higgs
masses which determine the weak scale.
Thus we demand
\begin{equation}
  \label{gluinomass}
  m_{gluino}
  \simeq \frac{\alpha_3}{4 \pi}
  \delta \frac{f^2}{16 \pi^2}
  \frac{\lambda_Z^4 \lambda_Y^\frac{3}{2}}{2 f^{1 \over 2}} \Lambda 
  \simeq (10^2 - 10^3) \, {\rm GeV} .
\end{equation}

Second, the gravitino mass $m_{3/2}$ should be less than about 1 GeV
to keep the mass degeneracy of the squarks and sleptons explaining the 
required suppression of FCNC \cite{FCNC}:%
\footnote{We assume that there exists an additional sector to set the
  cosmological constant vanishing. Then the squarks and sleptons in
  the SUSY standard model acquire soft SUSY-breaking masses of order
  $m_{3/2}$ due to supergravity effects.}
\begin{equation}
  \label{fcnc}
  m_{3/2} \simeq \frac{\langle F_Z \rangle}{\sqrt{3} M} \simeq
  \frac{\lambda_Z \Lambda^2}{\sqrt{3} M} \leq 1 \, {\rm GeV},
\end{equation}
where $M$ is the gravitational scale, that is, $M \simeq 2 \times
10^{18}$ GeV.

In terms of the parameters, $\lambda_Z \simeq f \simeq 1$, for
instance, we get
\begin{equation}
  \lambda_Y^{3 \over 2} \Lambda \simeq (10^6 - 10^7) \, {\rm GeV}
\end{equation}
from Eq.(\ref{gluinomass}).
By means of Eq.(\ref{fcnc}), this gives
\begin{equation}
  \lambda_Y
  \mbox{\raisebox{-0.7ex}{ $\stackrel{\displaystyle >}{\sim}$ }}
  10^{-2}, \ \Lambda
  \mbox{\raisebox{-0.7ex}{ $\stackrel{\displaystyle <}{\sim}$ }}
  10^9 \, {\rm GeV} .
\end{equation}
For $\lambda_Y \mbox{\raisebox{-0.7ex}{ $\stackrel{\displaystyle
      <}{\sim}$ }} 1$ we get
\begin{equation}
  10^{-2}
  \mbox{\raisebox{-0.7ex}{ $\stackrel{\displaystyle <}{\sim}$ }}
  \lambda_Y
  \mbox{\raisebox{-0.7ex}{ $\stackrel{\displaystyle <}{\sim}$ }}
  1, \
  10^6 \, {\rm GeV}
  \mbox{\raisebox{-0.7ex}{ $\stackrel{\displaystyle <}{\sim}$ }}
  \Lambda
  \mbox{\raisebox{-0.7ex}{ $\stackrel{\displaystyle <}{\sim}$ }}
  10^9 \, {\rm GeV} .
\end{equation}
The gravitino mass is given by
\begin{equation}
  m_{3/2} \simeq 1 \, {\rm keV} - 1 \, {\rm GeV} .
\end{equation}
In the case of $m_{3/2} = (1 - 100)$ keV, the reheating temperature
$T_R$ after inflation should be very low as $T_R
\mbox{\raisebox{-0.7ex}{ $\stackrel{\displaystyle <}{\sim}$ }} 10^2$
GeV to avoid the overclosure of the universe \cite{moroi}.
Thus, we must invoke a late-time baryogenesis like in the Affleck-Dine 
mechanism \cite{affleck}.

\section{Conclusion}
\label{conclusion}

In this final section, we deal with a few remaining aspects of our
model.

First we consider the Polonyi problem
on the singlet $Z$. 
The mass of $Z$ is given by
\begin{equation}
  m_Z \simeq \sqrt{\eta} \lambda_Z^3 \Lambda \simeq (10^6 - 10^9) \,
  {\rm GeV}.
\end{equation}
This shows that the singlet causes no cosmological problem
\cite{kawasaki}.

Second we argue a naturalness problem in the messenger sector.
The relevant superpotential is given by
\begin{equation}
  W = \lambda_Z (QQ) Z
  + \lambda_Y (QQ) Y - \frac{f}{3} Y^3 + (k_1 \bar{d} d +
  k_2 \bar{l} l) Y,
\end{equation}
where $(QQ)$ denotes the right-hand side in Eq.(\ref{SPsinglet}).
We can consistently impose U(1)$_R$ symmetry%
\footnote{This $R$ symmetry is anomalous
and there is no light $R$-axion.}
with the charges of $Q_i, Y, Z, d, \bar{d}, l, \bar{l}$ as $2/3$.
However, this $R$ symmetry allows additional terms such as $f_Z Z^3$.
In order for our SUSY-breaking vacuum to be (meta)stable, the coupling
$f_Z$ should be extremely small.%
\footnote{This may be natural in the sense of 't Hooft under a
  symmetry which imposes $\lambda_Y = f_Z = 0$.
  The fact that $\lambda_Y \gg |f_Z|$ may originate from the charge
  difference between the two terms $\lambda_Y (QQ) Y$ and $f_Z Z^3$.}

Third we comment on the $\mu$-problem.
Let us consider an interaction $h Y\bar{H} H$ as a direct source for
the $\mu$-term $\mu \bar{H} H$ in the superpotential:
\begin{equation}
  \mu = h \langle Y \rangle.
\end{equation}
Then we also obtain a soft mass term $B \bar{H} H$ in the potential
with
\begin{equation}
  B = h \langle F_Y \rangle.
\end{equation}
Together with Eq.(\ref{gauginomass}), these yield
\begin{equation}
  \mu m_{gluino} \simeq {\alpha_3 \over 4\pi} B,
\end{equation}
which implies that the value $B$ is too large.
Thus we should give up the direct generation of the $\mu$-term.
When we obtain an appropriate $\mu$-term by some way or others
\cite{dine}, the SU(2)$_L \:\times$ U(1)$_Y$ breaking may be induced
by radiative corrections.

We note that even the messenger singlet $Y$ may be unnecessary if an
appropriate mass terms for the messenger quarks and leptons $d,
\bar{d}, l, \bar{l}$ are generated like the $\mu$-term.
Then the SUSY breaking may be mediated by introducing a messenger
gauge interaction for the messenger quarks and leptons as well as
$Q_i$.

Finally we comment on hidden sector models of DSB with vector-like
gauge theories.
Since SUSY-breaking is communicated by gravity in hidden sector
models, we need to introduce no messenger sector (no fields such as
$Y, d, \bar{d}, l, \bar{l}$), which enables us to respect naturalness
thoroughly.
Sizable gaugino masses stem from terms of the form $(Z/M)W_\alpha
W^\alpha$, which is compatible with U$(1)_R$-charge assignments of $Z$
and $(QQ)$ as zero and two, respectively.
Then we get an effective superpotential
\begin{equation}
  W_{eff} = \lambda_Z \Lambda^2 Z (1 + {\cal O}({Z \over M})).
  \label{hidden}
\end{equation}
We expect a SUSY-breaking local minimum at $|Z/M| \ll 1$ for the
effective potential corresponding to Eq.(\ref{hidden}) in supergravity
\cite{yanagida} in view of section 2.
When $\Lambda \simeq 10^{11}$ GeV and $\lambda_Z \simeq 1$, the
gravitino mass turns out to be $m_{3/2} \simeq 10^3$ GeV (see
Eq.(\ref{fcnc})), which characterizes the SUSY-breaking scale in the
standard-model sector.
Note that $m_Z \simeq 10^{11}$ GeV in this case so that the singlet $Z$
causes no cosmological problem.
However, in contrast to the visible sector models, the FCNC problem is
not automatically resolved in the hidden sector models.

\newpage

\section*{\large Figure caption }

\begin{description}
\item[Fig.1:]
Typical Feynman diagrams generating $f$-dependent corrections to the
K\"ahler potential.
\end{description}

\newpage

\end{document}